\documentclass[11pt,twoside]{article}
\usepackage{asp2010}

\resetcounters

\begin{document}

\title{Affordable Digital Planetariums with WorldWide Telescope}
\author{Philip Rosenfield$^1$, Andrew Connolly$^1$, Jonathan Fay$^2$,Larry Carey$^1$,	
Conor Sayres$^1$, and Benjamin Tofflemire$^1$
\affil{$^1$Astronomy Department, University of Washington Box 351580, Seattle, WA 98195-1580}
\affil{$^2$Microsoft Research, One Microsoft Way, Redmond, WA 98052 U.S.}}

\begin{abstract}
Digital planetariums can provide a broader range of educational experiences than the more classical planetariums that use star-balls. This is because of their
ability to project images, content from current research and the 3D distribution of the stars and galaxies. While there are hundreds of planetariums in the country the reason that few of these are full digital is the cost. In collaboration with Microsoft Research (MSR) we have developed a way to digitize existing planetariums for approximately \$40,000 using software freely available. We describe here how off the shelf equipment, together with MSR's WorldWide Telescope client can provide a rich and truly interactive experience. This will enable students and the public to pan though multi-wavelength full-sky scientific data sets,  explore 3d visualizations of our Solar System (including trajectories of millions of minor planets), near-by stars, and the SDSS galaxy catalog.
\end{abstract}

\section{Introduction}

The 8-meter diameter planetarium at the University of Washington (UW) is primarily used for formal education as a part of our introductory astronomy classes. Graduate students also organize and coordinate informal educational presentations to K-12 students and astronomy clubs as well as programs for the general public. With an MS-8 star-ball projector and three VGA projectors, the range of content of the presentations was limited by the lack of visualizations or at least constrained by the canned videos available. WorldWide Telescope\footnote{http://www.worldwidetelescope.org} (WWT) as a full dome digital projection system effectively eliminates the lack of visualizations, and with this software available to students and presenters every user can create their own tours and presentations; removing the constraints of pre-made media.

We present our methods to digitize the UW planetarium on a tight budget i.e., \$40,000 total (\$10,000 for computers and \$30,000 for projectors). In section \ref{proj} we discuss our choice of projectors and the need for conversion lenses and mounts, in section \ref{comp} we explain the decisions involved in choosing the computer system. With the projectors and computers nominally set up, we explain the alignment procedure in section \ref{alignm} and in section \ref{conc} we conclude by describing the expected integration of this digital
dome within the educational curriculum.

\section{Choosing Projectors}
\label{proj}
\begin{figure}[p]
\includegraphics[width=.5\textwidth]{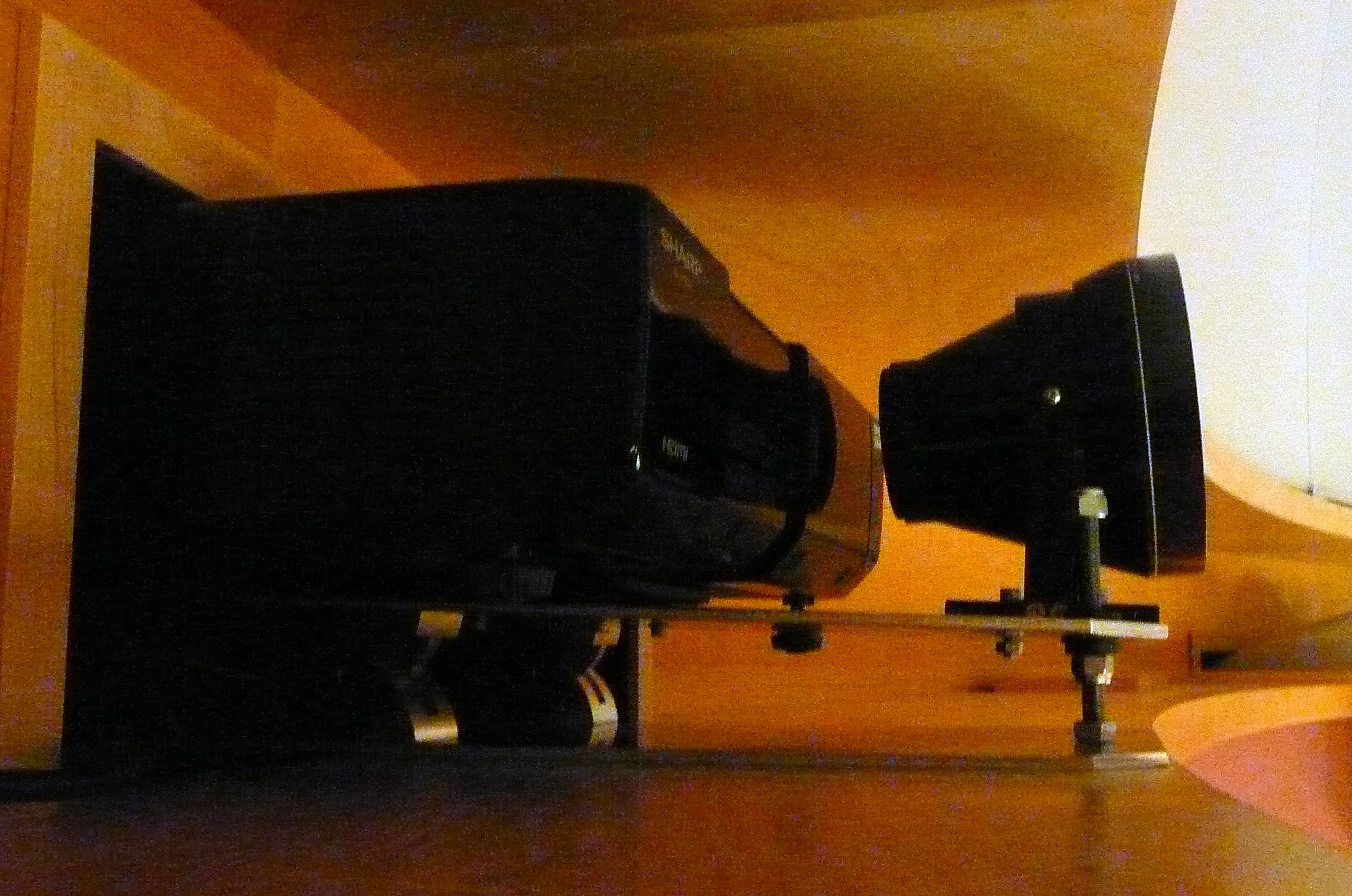}
\caption{On of the four projectors that illuminate the perimeter of the planetarium dome stationed on a mount with a Navitar conversion lens}
\label{venus}
\end{figure}

\begin{figure}[p]
\includegraphics[width=.5\textwidth]{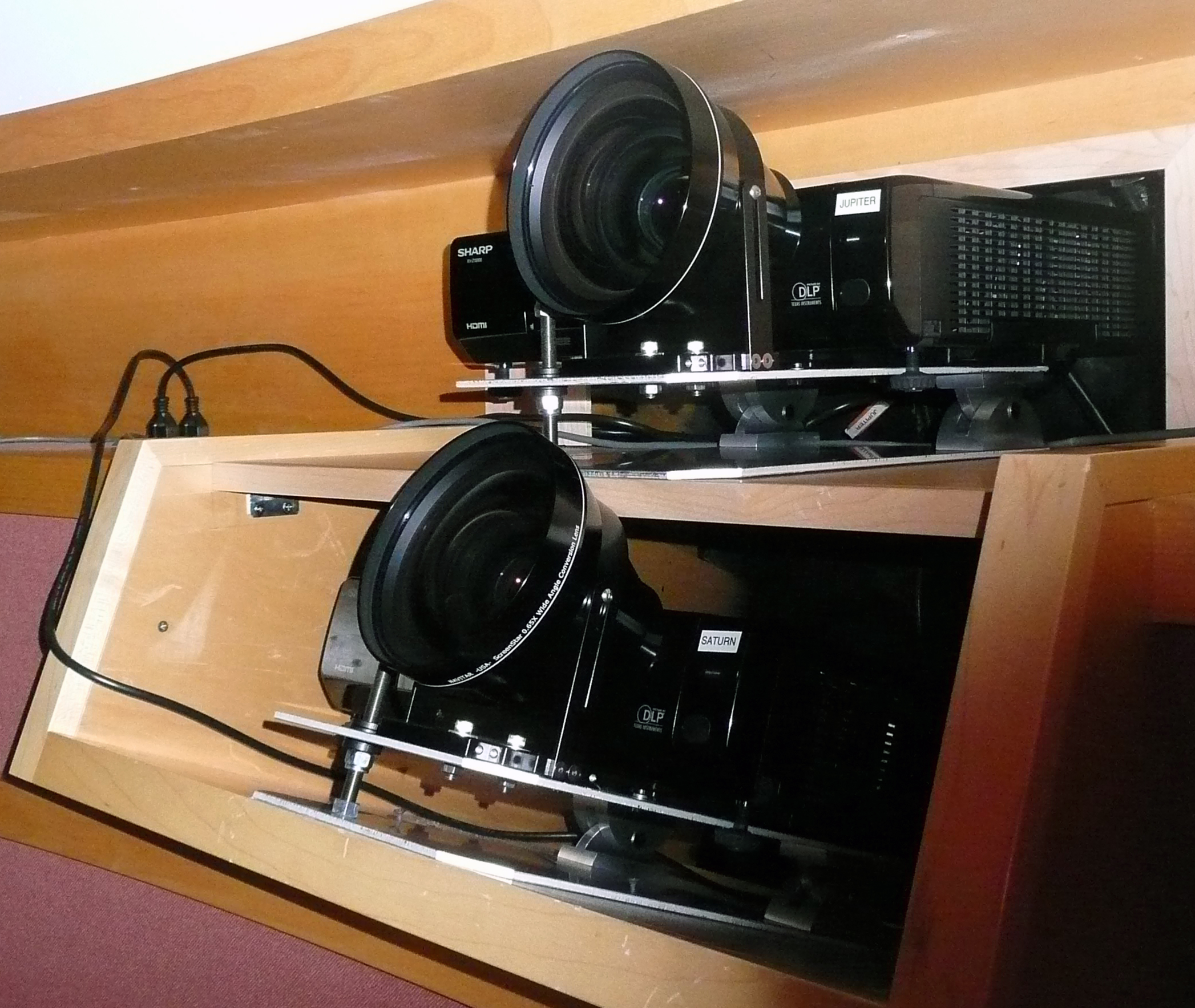}
\caption{A perimeter projector and a cap projector below mounted with conversion lenses. The cap projector is replacing the position of a old projector. }
\label{jupiter}
\end{figure}

With the MS-8 projector in the middle of our dome and a shelf already built into the perimeter of the planetarium about a foot under the spring-line (see Figure \ref{venus}), we had a natural choice to place four projectors around the perimeter, with two projectors aimed to the top of the dome (see Figure \ref{jupiter}. We considered alternative designs including placing the projectors at ground-level in the center of the dome, and a single projector for the dome cap. Each of these scenarios had significant drawbacks.
These include: space limitations that restrict the number of projectors we can mount in the center of the dome, a concern that the the projectors would need to be shielded to block the projector beams from the eyes of the patrons, and few projectors in our price range with the features we desire are designed to be positioned at a steep angle. Ignoring this final limitation could at worst burn out a projector (since the fans would not be functioning properly) or at best void any warrantee.

The goals in choosing the projectors were to provide a rich and immersive environment with as high a resolution as possible. As projectors are the primary cost of this upgrade we sought machines that would have longevity, good dynamic range, excellent black response and that would not be obsolete on a timescale
of a few years. We limited our search such that our projectors budget would be under \$30,000, we would need the maximum cost of a projector to be \$5,000. In the summer of 2009, it seemed impossible to find a projector at that cost such that when we placed four of them around the circumference of our planetarium, they filled the entire width including some extra space for the projected images to overlap. In fact, there seemed to be a jump in projector prices from ``Home Theater'' (\$1,000-\$4,000) to professional grade ($>\$15,000$)\footnote{We used http://www.projectorcentral.com/}. Limited by Home Theater projectors, while ensuring the richest immersive experience that would also stay ``cutting edge'' for the longest time, we limited our search to projectors with the highest resolution available in 2009, 1920x1080p. This resolution usually implied an aspect ratio (width of projected image $W$ to height of projected image) of 16:9.

In a dome of a given diameter, $d$, each projector must fill a width of $W > \pi d /4$. In the terms used in the projector industry, the Home Theater projectors have throw ratios (distance to screen, $D$, to width of image) of about 1.8-3. So, as an estimate of the throw ratio necessary for a dome environment, our projectors would be placed about the diameter of the dome away from the image. This makes the throw ratio $ D/W < 1.27 D/d$, and by our approximation, $D/d \approx 1$. This means the throw ratio would need to be 1.27 if there were no overlap, and less than 1.27 to have any overlap.

We overcame this ``short throw'' problem by using conversion lenses. Navitar\footnote{http://www.navitar.com}, a company that makes conversion lenses, makes the ScreenSaver 0.65 Conversion Lenses listed at \$2,500 each, thus defining the same cost as the upper limit to the cost of a projector on our budget. With a 0.65 conversion lens, the projected image width will be increased by a factor of 1.5 and we can move the projector closer to the screen by about 33\%. In other words the throw ratio will change to $D^{'}/W^{'} = 1.5D/0.65W =  2.3 D/W $. With our demand of a $D/W < 1.27$, we need to find a projector with a throw ratio of less than 2.9, an easy feat in the Home Theater market.

There are many Home Theater projectors available for under \$2,500 that have the resolution we require, and with the constraint of the throw ratio relaxed, the next feature we looked into was the contrast ratio. For our purposes, the contrast ratio is a red herring. It is supposed to compare the brightest pixel to the darkest pixel and thus giving the consumer an idea of how dark the projection when the pixel is defined to be black. With some projectors having automatically adjusting irises (aka a dynamic iris, an iris that will block light instead of projecting dark), the ratio is not comparable from projector to projector. The contrast ratio becomes more of a distraction since it takes the control away from us to make what we want dim or bright. Since we wont be using the dynamic iris features on the projector, we looked for low lumen projectors with good dark levels with the assumption our audience's eyes will adjust to a very dark room. This lead us to look at the projection type, that is Liquid Crystal Display (LCD) or TI's Digital Light Processing (DLP)\footnote{http://www.dlp.com/}. LCD projectors have less pixellation issues but are known to have hot or dead pixels (when LCD panels have defective transistors), while DLP projectors tend to do better with dark levels (since tiny mirrors point the light away from the projection screen), thus we chose to limit ourselves to a DLP projector with around 1600 lumen brightness. To nail down the brightness levels, we contacted Avidex\footnote{http://www.avidexav.com/}, a Seattle based Audio/Visual design company, to compare different projector models. Based on these tests we selected the Sharp XV-Z15000\footnote{http://www.sharpusa.com/ForHome/HomeEntertainment/FrontProjectors/XVZ15000} but we believe that the appropriate choice for a given
planetarium will depend on the dome surface and the configuration of the planetarium itself.

\subsection{Projector Mounts and their Placement in the Planetarium}
After picking out the locations of the projectors in the planetarium, we needed to allow small amounts of freedom in rotation and inclination so we can control the overlap areas but lock down the focus. The need for conversion lenses added extra complications. We needed to optically align the lenses and projectors and find rigid placements for the mounted systems. To do this, mounts were designed by one of us (LC), an optical engineer with the UW Telescope Engineering Group.  

Once the physical location (x, y and z) of the projectors had been defined, the remaining two degrees of freedom that required adjustability for final positioning the projector were azimuthal rotation and rotation in altitude. Azimuthal rotation is handled using the two fasteners that attach the base of the adjustable mount to the selected mounting surface of the planetarium. The mounting base plate is designed with one mounting hole that is bolted down to the mounting surface in the planetarium at the desired ``x, y, and z'' location for the projector, providing a pivot point for the base plate and plane for azimuthal rotation. The second fastener fits into a curved slot that is centered on the pivot point, allowing for azimuthal rotation of the mount and projector assembly around the first fastener. Rotation in altitude is controlled by hinging, on a horizontal axis, the mounting plate on which the projector sits, relative to the base plate that is mounted to the planetarium's mounting surface. The altitude adjustment is controlled using spherical washers and nuts on a fine thread screw that can fix the projector mounting plate in the desired position using lock nuts. The plate on which the projector is mounted also provides the mounting surface for the lens (mounted in front of the projector) that is needed to re-size the projector image for the planetarium dome. The custom designed hinges and mounting plates were manufactured and assembled by UW Astronomy students (BT and CS) in the UW Physics Student Machine Shop (for engineering drawings please contact the authors).

\section{Choosing Computers}
\label{comp}
Computer technology changes faster than the speed of conference proceedings so we will limit the amount of detail in this section. Our goal was to maximize the video display possible while minimizing the cost of the system. We chose to use seven computers, six of them attached to the projectors in client mode and the seventh, the master mode computer. For our video card, we chose NVIDIA's GTX275. The seven computers all have Windows 7 OS with a 1 TB hard drive and 6 GB RAM. To communicate from one computer to the another (for projector alignment, software instillation, and running WWT) an ethernet switch is required and a Virtual Network Computing (VNC) program is recommended. Windows 7 has a home network setup with remote connections possible, making it easy to install a program from the master computer on a client, but the software logs the user out of the client computer, while a VNC program will allow the master user to see their actions on the client monitor (or in this case, projected on the dome). Another consideration when purchasing multiple computers is the option for ``Wake on LAN'' in the system BIOS of the motherboard chosen. This would allow the master computer to turn on the other machines instead of having to physically switch on each computer at every use.

\section{Alignment}
\label{alignm}
Once the projectors, lenses, and mounts are in place, the software must know which pixels go where, how they are distorted, and which pixels are in the overlapped sections of the projection. Projection Designer\footnote{http://orihalcon.jp/projdesigner/}, a manual set up tool to correct for distortion and edge blending, is a free and open source way to align the projectors in your planetarium or any multi-projector environment. It is not a trivial procedure and takes some time, we found that it is easiest with more than one person. Starting with the projector aimed at the ``sweet spot'' the place on the dome that will be the apex of any traveling-like movement, we aligned a projected full dome grid to points and lines existing on our dome or projected from our MS-8 star-ball. The output of Projection Designer is a set of blend and distortion maps that WWT has been designed to read. At this point, see the WWT documentation for instructions on incorporating Projection Designer's output files\footnote{http://www.worldwidetelescope.org/Docs/worldwidetelescopeplanetarium.html}.

\section{Conclusions}
\label{conc}

We've shown how we've digitized the University of Washington's planetarium with off-the-shelf equipment and WWT software. Table \ref{t-costs} contains a summary of our costs. We will be incorporating this new technology into formal and informal education at all levels. 

In addition to the planetarium experience we have all grown up with, WWT can currently project more than thirteen all-sky data sets onto the night sky covering nearly every detectable band of the electromagnetic spectrum even allowing cross fading from one to another.

WWT also has 3-D interactive visualization with the Hipparcos catalogue and Sloan Galaxy Map. With a click of a mouse or a twist of joy stick or XBOX controller, the presenter or the student from home can fly through the observed universe: look at the planets orbit the Sun from the vantage point of Pluto, see how the night sky appears from a star far from the Sun, watch a swarm of millions of minor planets in their trajectories around the sun, and zoom through the Coma Cluster of galaxies. This is a huge visualization upgrade from a geocentric star-ball.

For upper level undergraduate work, WWT can make real time calls to astronomical databases to access research grade data on ADS, Simbad, SEDS, and NED, it can download existing FITS or JPG image files, and even import FITS files that students in an observing class would take with a telescope and place them correctly in the sky. At the UW, a 16" telescope for undergraduate projects is very near our planetarium, and future capabilities would drive it from the planetarium displaying the raw image taken on the dome. Students are to search the Virtual Observatory registry and overlay data or output .xml files to be plotted or analyzed in a myriad of ways. For general information they can simply right click on the object in question and look it up directly in Wikipedia.

Instructors wishing to duplicate or save their own classroom planetarium shows or astronomy public outreach talks can create a tour to run an automated planetarium show that would record their own voice, incorporate images and music from external sources, and write text as if it were a dome filled PowerPoint presentation. The tour could then be saved to disc for the instructor to use each quarter, for a TA to play for the class, or to upload for the WWT online community. A slew of guided tours are already available, created by astrophysicists from the Adler Planetarium in Chicago, the Smithsonian Center for Astrophysics, the Space Telescope Science Institute and more.

With WWT in a full digital dome, all levels of undergraduates in astronomy could begin classroom labs in the planetarium and finish their assignments on their home computers or laptops. Graduate students and advanced undergraduate students will have a unique opportunity to create truly interactive shows for the public. We've outlined how we put it together costing less than \$40,000.

\begin{table}
\label{t-costs}
\caption{Cost Breakdown on Materials (all costs rounded up)}
\smallskip
{\small
\begin{tabular}{lr}
\tableline
\noalign{\smallskip}
Item & Cost \\ 
\noalign{\smallskip}
\tableline
\noalign{\smallskip}
Computers (White boxes) & \$15,000 \\
Navitar ScreenStar 0.65 Conversion lenses \& Sharp XV-Z15000 Projectors & \$22,000 \\
HDMI Cables & \$350 \\
Misc (Ethernet switch, security locks) &   \$200 \\
\textbf{\textit{Total}} & \textbf{\textit {\$38,000}} \\
\noalign{\smallskip}
\tableline
\end{tabular}
}
\end{table}

\acknowledgements The authors with to thank MSR and College of Arts and Science at the University of Washington.


\end{document}